\documentclass[a4paper]{article}
\usepackage{multirow}
\usepackage{INTERSPEECH2021}
\usepackage{xcolor}
\usepackage{hyperref}
\usepackage{graphicx}
\usepackage{enumitem}

\makeatletter
\newcommand{\printfnsymbol}[1]{%
  \textsuperscript{\@fnsymbol{#1}}%
}
\makeatother

\title{Human Listening and Live Captioning:\\Multi-Task Training for Speech Enhancement}
\name{Sefik Emre Eskimez\thanks{\printfnsymbol{1}Equal contribution}\printfnsymbol{1}, Xiaofei Wang\printfnsymbol{1}, Min Tang, Hemin Yang, Zirun Zhu,\\Zhuo Chen, Huaming Wang, Takuya Yoshioka}

\address{Microsoft, One Microsoft Way, Redmond, WA, USA}
\email{\{seeskime, xiaofewa, mintang, heyang, zirzhu, zhuc, huawang, tayoshio\}@microsoft.com}

\begin{document}

\maketitle
\begin{abstract}
    With the surge of online meetings, it has become more critical than ever to provide high-quality speech audio and live captioning under various noise conditions. However, most monaural speech enhancement (SE) models introduce processing artifacts and thus degrade the performance of downstream tasks, including automatic speech recognition (ASR). This paper proposes a multi-task training framework to make the SE models unharmful to ASR. Because most ASR training samples do not have corresponding clean signal references, we alternately perform two model update steps called SE-step and ASR-step. The SE-step uses clean and noisy signal pairs and a signal-based loss function. The ASR-step applies a pre-trained ASR model to training signals enhanced with the SE model. A cross-entropy loss between the ASR output and reference transcriptions is calculated to update the SE model parameters. Experimental results with realistic large-scale settings using ASR models trained on 75,000-hour data show that the proposed framework improves the word error rate for the SE output by 11.82\% with little compromise in the SE quality.  
    Performance analysis is also carried out by changing the ASR model, the data used for the ASR-step, and the schedule of the two update steps. 
\end{abstract}

\section{Introduction}
Due to the need for social distancing led by the COVID-19 pandemic, people and organizations have had to rely on digital technologies to stay connected and work remotely~\cite{pandey2020impact}. This has resulted in a surge in the usage of online conferencing tools.
Most organizations rely on these tools to conduct day-to-day business while people rely on them to connect with their family members and friends in these challenging times. 
As a result, it is becoming more important than ever to provide high-quality speech audio under various household noise conditions.

In recent years, deep neural networks have shown great potential for single-channel speech enhancement (SE) (or noise-suppression)~\cite{weninger2015speech,hu2020dccrn,yin2020phasen,kinoshita2020improving,choi2018phaseaware,tang2020joint,isik2020poconet}. Although these models substantially remove background noise, most of them degrade the performance of downstream tasks such as automatic speech recognition (ASR) performance significantly,  as modern commercial multi-condition trained ASR systems can usually recognize original noisy speech~\cite{kinoshita2020improving} well and the SE models introduces unseen distortions that are particularly harmful to ASR. When both live captioning and high-quality audio are needed, a common solution is for the local client to send both an enhanced signal for communication and an unaltered signal for transcription. Clearly, there is a benefit of creating a speech enhancement system that can improve speech quality without compromising the ASR accuracy.  

In this paper, we propose a framework for optimizing speech enhancement models for both communication and transcription quality by leveraging pre-trained ASR models. The framework aims to build an SE model that achieves superior ASR performance while retaining the same speech quality as an SE model trained solely for SE objectives. 

Our training framework alternately performs two steps: SE-step and ASR-step. In the SE-step, the model is trained with parallel data created by artificially mixing clean speech with noise files as with conventional supervised SE training. For the ASR-step, we use a realistic ASR model trained on a large amount of data. 
The training audio is passed through the SE network and fed to the ASR network to calculate an ASR-based loss. 
In both steps, only the SE network is updated while the ASR model parameters remain unchanged. This training scheme allows us to leverage real noisy recordings which do not have the corresponding clean speech signals to optimize the SE network. We evaluate our framework by using both real and synthetic data under various conditions with respect to signal-to-noise ratios (SNRs), types of noise, and recording scenarios. The experimental results show that our proposed framework improves the word error rate (WER) by 11.82\% over the state-of-the-art causal DCCRN model~\cite{hu2020dccrn} for 
real recordings. Besides, we conduct ablation studies by changing the ASR model, the data for the ASR-step, and the schedule of the two training steps. 
We use different ASR systems for the training and evaluation to show the generalization capability of the proposed framework. 

\section{Related Work}

\textbf{Speech Enhancement:} There are regression-based and masking-based approaches for speech enhancement using neural network models~\cite{zhao2018convolutional,hu2020dccrn}. Regression-based approaches try to predict a clean speech signal or its time-frequency representation (TF) from the noisy speech input. Masking-based methods try to estimate a TF mask from the noisy input and apply the predicted mask to the same input to obtain the clean signal. 


Various architectures, along with many objective functions, were proposed for solving the SE problem. We focus on the most recent and promising work. 
Choi et al.~\cite{choi2018phaseaware} proposed a deep complex U-net (DCUNET) that uses real-valued convolution operations on the real and imaginary parts of the input speech STFT. They took advantages of the U-net structure and deep complex networks~\cite{deepcomplex}, which was shown to be useful for SE~\cite{pascual2017segan}. They employed a scale-invariant signal-to-noise ratio (SI-SNR) calculated in the time domain to optimize their model parameters. Hu et al.~\cite{hu2020dccrn} built upon the DCUNET and introduced a deep complex convolution recurrent network (DCCRN). They introduced complex long-short term memory (LSTM) layers in the bottleneck layer and complex batch normalization. The resulting model was significantly faster and had fewer parameters than DCUNET.


Kinoshita et al.~\cite{kinoshita2020improving} considered using a convolutional time-domain audio separation network (Conv-TasNet) for SE and proposed Denoising-TasNet, which directly processes a speech waveform with 1D convolutions. Their results show that the time-domain SE approach achieved a relative WER improvement of more than 30\% 
for a robust ASR back-end. However, we could not observe WER improvements in our preliminary evaluation with time-domain approaches using a production-grade ASR back-end trained from both clean and noisy audio. Hence, we do not consider this approach in this work. 

\textbf{ASR Multi-task Training: }
There are prior studies that employ joint or multi-task training for the front-end and back-end~\cite{seltzer2004likelihood,chen2015speech,giri2015improving,pironkov2016multi,xiao2016deep,meng2017deep,sainath2017multichannel,minhua2019frequency,subramanian2019speech,watanabe20202020,wang2020exploring}. However, they concerned only about the ASR accuracy and did not pay close attention to the SE quality nor analyzed the trade-off between the two tasks. 
In contrast, we focus on the front-end and optimize it for both tasks. 

\section{Method}
This section elaborates on our proposed framework. 
We first present the proposed multi-task training framework using SE and ASR. 
In the subsections that follow, we describe the pieces that constitute the proposed framework, i.e., 
our SE model as well as the SE and ASR loss functions to be used. 

\begin{figure}[t]
  \centering
  \includegraphics[width=\columnwidth]{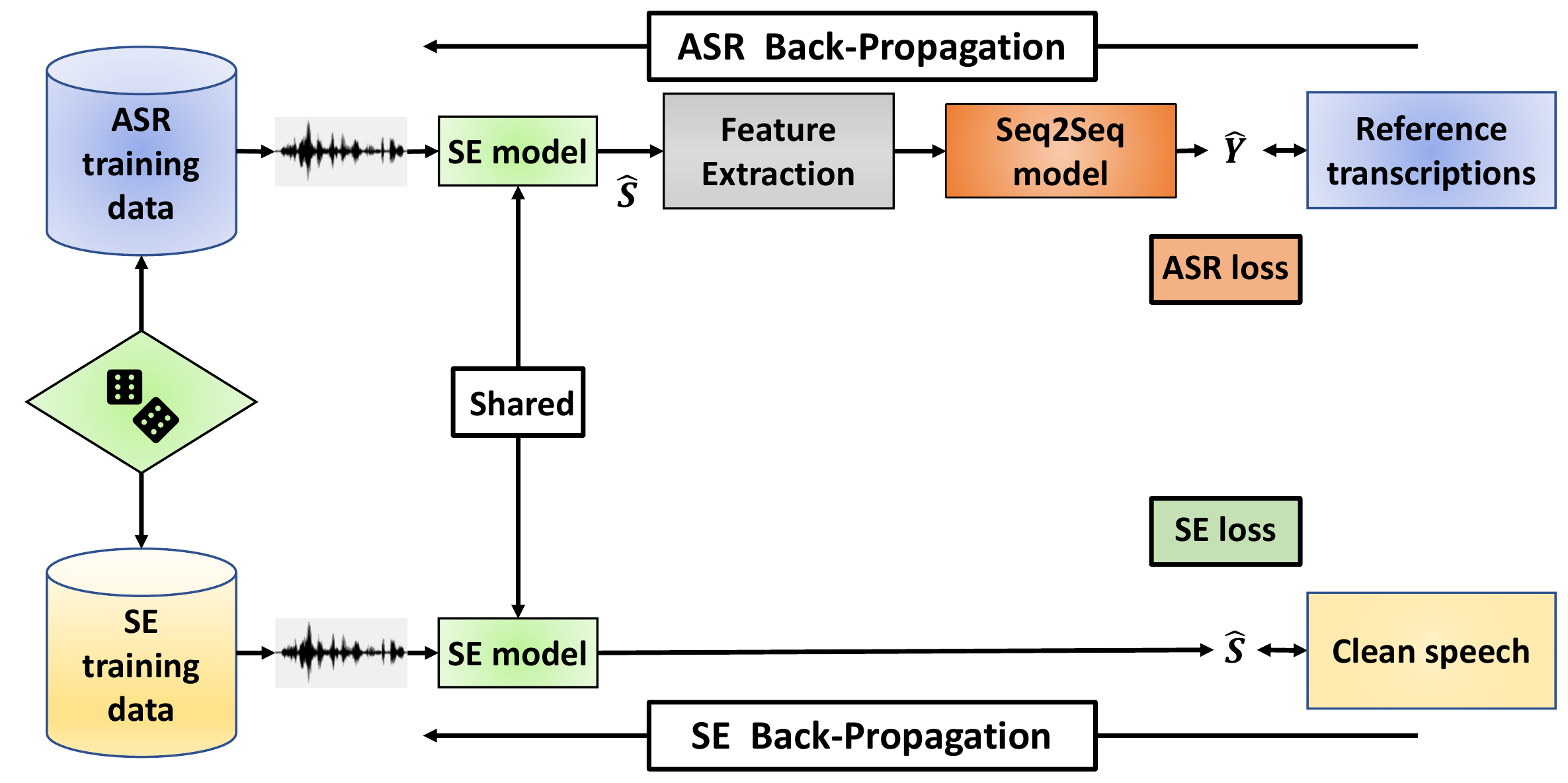}
  \vspace{-0.7cm}
  \caption{Proposed multi-task training framework. $\hat{S}$ and $\hat{Y}$ denote enhanced signal and predicted transcription, respectively.}
  \vspace{-0.5cm}
  \label{fig:sysoverview}
\end{figure}


\subsection{Multi-Task Training} \label{sec:multi-task}
\vspace{-0.3em}
The idea behind the proposed method is to optimize an SE model for both noise suppression and ASR. 
Multi-task training is usually performed by using training samples that have supervision signals (i.e., reference signals and word labels) for all the tasks to be considered. 

We consider a scenario where each training sample has a supervision signal for only either SE or ASR. 
This is because, in practice, most of the ASR datasets do not contain training samples with reference clean speech signals that could be used for supervised SE training. Also, some SE datasets do not come with human transcriptions. 


To cope with this scenario, we update the SE model parameters by alternately performing the following two steps: 
\begin{description}[style=unboxed,leftmargin=0cm]
    \item[SE-step:] We mix clean speech and noise samples on the fly and use the noisy and clean speech pairs to evaluate a signal difference-based loss function. The SE model parameter gradients with respect to this loss function are computed to update the model parameters. 
    \item[ASR-step:] Noisy training samples in a mini-batch are fed to the SE network. The generated enhanced signals are input to the ASR network. We evaluate a loss function by comparing the ASR model output and the reference transcriptions. The loss is back-propagated all the way down to the SE network, and only the SE model parameters are updated. 
    This is because our objective is to find SE model parameter values that would work for existing well-trained ASR systems, and therefore we do not want the ASR network to adapt to the characteristics of the SE model. 
\end{description}

Figure~\ref{fig:sysoverview} shows the diagram of the proposed framework. 
The two-step approach allows us to take advantage of the real noisy speech samples that only have reference transcriptions for the SE model training. 
At each training iteration, the update step to be used is chosen randomly from a Bernoulli distribution. 
We refer to the probability of choosing the SE-step as the ``SE-step probability''. 
Before performing the multi-task training, the SE model parameters are pre-trained on the SE training dataset. 


\subsection{Speech Enhancement Model} \label{sec:se_model}
\vspace{-0.3em}
We employ DCCRN~\cite{hu2020dccrn} as our front-end model because it achieved the best SE performance in our preliminary test, while we also provide results using DCUNET to examine the generalization capability of the proposed framework. 
DCCRN applies an encoder-decoder architecture with two LSTM layers in between.  Instead of using conventional 2D convolutional/deconvolutional layers (conv2D/deconv2D), DCCRN builds on the DCUNET~\cite{choi2018phaseaware} that employs complex conv2D/deconv2D followed by complex batch normalization in the encoder and decoder blocks. Besides, DCCRN employs complex LSTM layers instead of conventional LSTM layers. Furthermore, it contains U-Net style skip connections (concatenation) from the encoder to the decoder as in DCUNET. 

The DCCRN model takes the real and imaginary parts of a noisy spectrogram as input. It estimates a complex ratio mask (CRM) and applies it to the noisy speech. The masked signal is converted back to the time domain with ISTFT. The CRM can be defined as follows: 
\begin{equation}\label{stargan}
\begin{split}
CRM = \frac{Y_{r}S_{r}+Y_{i}S_{i}}{Y^{2}_r+Y^{2}_i} + j\frac{Y_{r}S_{i}-Y_{i}S_{r}}{Y^{2}_r+Y^{2}_i}
\end{split}, 
\end{equation}
where $Y_r$ and $Y_i$ denote the real and imaginary parts of the noisy spectrogram, respectively while $S_r$ and $S_i$ denote the real and imaginary parts of the clean spectrogram, respectively. 

We consider both non-causal and causal DCCRN configurations. The non-causal model can look ahead and employs bidirectional LSTM layers. The causal model can only process the current and previous frames and employs unidirectional LSTM layers and causal padding for conv2D/deconv2D layers. The latter is suitable for most applications. 
For further details of DCCRN, we refer the reader to \cite{hu2020dccrn}. 

\begin{table*}[ht!]
\centering
\caption{Evaluation results for seed SE models and their multi-task (MT) trained versions. SE-step probability for MT training was set at $0.5$ except for the Best WER MT case, where it was set at $0.0$.}
\vspace{-0.3cm}
\label{tab:main_results}
\begin{tabular}{l|ccccc|cc}
\hline \hline
\multicolumn{1}{c|}{\multirow{2}{*}{\textbf{System}}} & \multicolumn{5}{c|}{\textbf{Simulation Data}}         & \multicolumn{2}{c}{\textbf{Real Data}} \\
\multicolumn{1}{c|}{}                                 & \textbf{WER} & \textbf{pMOS} & \textbf{PESQ} & \textbf{SDR} & \textbf{STOI} & \textbf{WER}      & \textbf{pMOS}      \\ \hline
Noisy Speech (No enhancement)                         & 24.02        & 2.81          & 1.42          & 5.02         & 0.82          & 19.47             & 2.91               \\ \hline 
DCUNet-causal-seed                                    & 26.36        & 3.54          & 2.23          & 11.51        & 0.88          & 23.28             & 3.18               \\
DCUNet-causal-MT                                      & 25.41        & 3.51          & 2.23          & 11.30        & 0.88          & 21.24             & 3.16               \\ \hline 
DCCRN-non-causal-seed                                 & 22.33        & 3.74          & 2.57          & 13.02        & 0.91          & 20.63             & 3.33               \\
DCCRN-non-causal-MT                                   & 21.92        & 3.69          & 2.59          & 12.90	    & 0.91          & 19.33             & 3.30               \\ \hline 
DCCRN-causal-seed                                     & 25.20        & 3.58          & 2.36          & 12.07        & 0.89          & 22.84             & 3.20               \\
DCCRN-causal-MT (Best WER Model)                      & 22.31        & 3.19          & 1.90          & 7.82         & 0.86          & 18.98             & 3.09               \\
DCCRN-causal-MT                                        & 23.34        & 3.48          & 2.31          & 11.70        & 0.89          & 20.14             & 3.17               \\ \hline
\end{tabular}
\vspace{-0.4cm}
\end{table*}

\subsection{Loss Function for SE-Step} \label{sec:se_loss}
\vspace{-0.3em}
For the loss function of the SE-step and the SE model pre-training, we use the PHASEN loss function~\cite{yin2020phasen}. 
It outperformed alternative loss functions, such as SI-SNR and power-spectrogram mean squared errors, in our preliminary tests. 

The PHASEN loss comprises two parts: amplitude $L_a$ and phase-aware $L_p$ losses. The definition is as follows:
\begin{equation}\label{phasenloss}
\resizebox{.9\hsize}{!}{$\mathcal{L} =  
\left||S|^p-|\hat{S}|^p\right|^2 + \left||S|^pe^{j\varphi (S)}-|\hat{S}|^pe^{j\varphi (\hat{S})}\right|^2$}, 
\end{equation}
where $S$ and $\hat{S}$ are the estimated and reference (i.e., clean) spectrograms, respectively. Hyper-parameter $p$ is a spectral compression factor and is set to 0.3. Operator $\varphi$ calculates the argument of a complex number. 

\subsection{Loss Function for ASR-Step} \label{sec:asr_model}
\vspace{-0.3em}
Our back-end model used for the ASR-step training 
is based on a sequence-to-sequence (Seq2Seq) model using an attention-based encoder-decoder structure \cite{wang2020exploring,chan2016listen}. We use the Seq2Seq model because it is simpler than a hybrid ASR system and thus facilitates multi-task training. 
The model estimates text sequence $C = (c_1, c_2, \cdots)$ from STFT $X$ that is generated by the SE model. 
We integrate STFT, log-mel filterbank energy extraction, and global mean-variance normalization into the ASR network to allow the gradients to pass through to the SE model.

In the ASR-step, the SE model parameters are updated to minimize the cross entropy loss between $C$ and reference label $R=\{r_1,...,r_N, r_{N+1}=\langle eos\rangle\}$, where $N$ is the length of the reference sequence $R$. Special symbol $\langle eos\rangle$ indicates the sequence end. 
See Sections 2.3 and 4.4 of \cite{wang2020exploring} for our Seq2Seq ASR model.

\section{Experiments}

We conducted ASR and SE experiments to evaluate the proposed multi-task training framework under realistic settings. 

\subsection{Datasets}
\vspace{-0.3em}
\textbf{Training Data for SE:}
\label{sec:train_data_se}
We utilized a large-scale and high-quality simulated dataset described in~\cite{braun2021efficient}, which includes around 1,000 hours of paired speech samples\footnote{We thank Sebastian Braun, Hannes Gamper, Chandan K.A. Reddy, and Ivan Tashev from Microsoft Research for providing us with the pre-mixed speech enhancement dataset and the pMOS tool.}. As a clean speech corpus, the dataset collects 544 hours of speech recordings with high mean opinion score (MOS) values from the LibriVox corpus~\cite{kearns2014librivox}. The mixtures are created using 247 hours of non-stationary noise recordings from the Audioset+Freesound~\cite{gemmeke2017audio,fonseca2017freesound} (187 hours), internal noise recordings (65 hours), and colored stationary noise (1 hour) as noise sources. In addition, the clean speech in each mixture is convolved with an acoustic room impulse response (RIR) sampled from 7,000 measured and simulated responses. See \cite{braun2021efficient} for details of this dataset. The data are available publicly, except for the 65 hours of the internal noise recordings\footnote{\url{https://github.com/microsoft/DNS-Challenge}}.

\noindent\textbf{Training Data for ASR: }
\label{sec:train_data_asr}
We trained our Seq2Seq ASR model based on 
64 million anonymized and transcribed English utterances, totaling 75K hours.

\noindent\textbf{Multi-task training Data: }
We used different training data for the SE and ASR-steps as described earlier. For the ASR-step, we used a subset of 75K-hour transcribed data. Section \ref{sec:in_out_domain_data} explores considerations for selecting the subset, such as 
in-domain vs. out-of-domain, and including vs. excluding simulated data. 
Note that the 75K-hour data for the ASR-step also contained augmented/simulated data and that these simulated data were different from the SE training data. 

\noindent\textbf{Evaluation Data: }
We used both simulated and real test data. 
The simulated test set comprised 60 hours of simulated audio with SNRs ranging from -10 dB to 30 dB. 
The simulation was performed as with 
the training data while clean speech signals were taken from LibriSpeech train-clean-100~\cite{LibriSpeech} and convolved with RIRs generated by using different configurations. We added both Gaussian and non-stationary noise. 
The latter was generated by convolving noise recordings from SoundBible+Freesound (10 hours in total) with simulated RIRs.

For the real test data, we employed two sets of noisy audio. The first set consisted of 18 hours of data recorded in an acoustically configurable audio lab where high-fidelity spatial noise sounds were played back from eight loudspeakers. The second set comprised 18 hours of meeting recordings. This test set contained various types of natural noise sounds. Furthermore, we included a clean speech test set consisting of 7803 words to measure the distortion introduced by the SE model.



\begin{figure*}[t!]
  \centering 
  \centerline{\includegraphics[width=\textwidth]{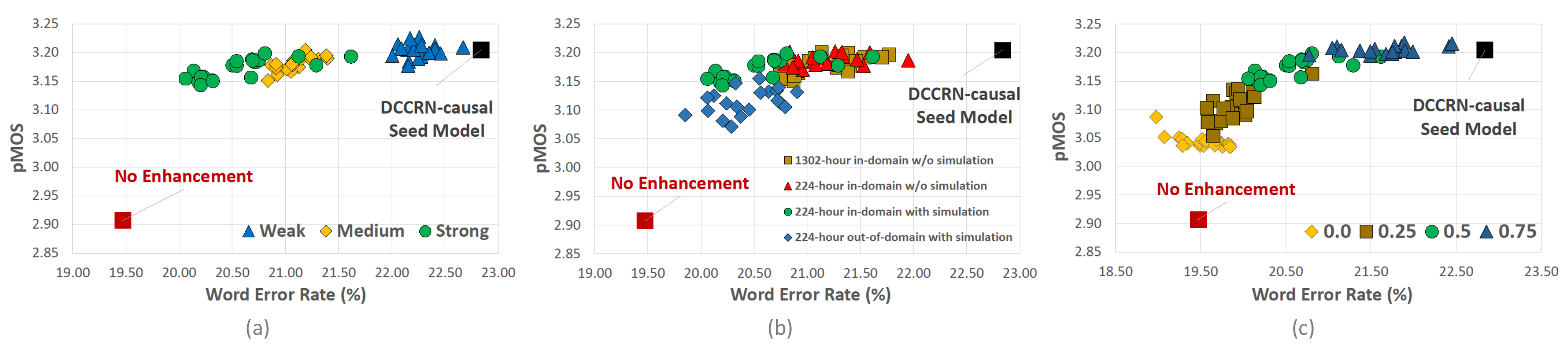}}
\vspace{-0.4cm}
\caption{(a) Performance comparison by using different ASR back-end models. ``Strong,'' ``Medium,'' and ``Weak'' in different colors represent three Seq2Seq models that have different ASR accuracies for the validation set. Each colored dot represents a checkpoint obtained every 5K iterations from multi-task training. (b) Impact of different ASR-step training datasets. Note that they are subsets of the ASR back-end model training data. ``In-domain'' means the training data is within a similar application scenario with the evaluation set. (c) Impact of SE-step probability on pMOS and WER.}
\vspace{-0.4cm}
  \label{fig:ablation}
\end{figure*}

\subsection{Implementation Details}
\vspace{-0.3em}
Our framework was implemented in PyTorch~\cite{paszke2019pytorch}. 
The seed (i.e., pre-trained) SE model was trained for 50 epochs with a batch size of 96 using 4 NVIDIA V100 GPUs.

In practice, we cannot fine-tune the SE model for every single back-end, which varies from time to time. Therefore, we used different back-end ASR models for training and evaluation, respectively. For ASR evaluation, a high-performance online hybrid ASR model was employed \cite{li2020high,li2019improving}. For multi-task training, the input feature for the offline Seq2Seq ASR model was 240-dimension log mel filterbanks, stacked by 3 frames with each frame having 10 msec. Global mean and variance normalization was applied before feeding the features to the ASR model encoder. We used 32K mixed-unit with $\langle space\rangle$ symbol between words as recognition units \cite{li2018advancing}. Label-smoothed cross-entropy loss \cite{chorowski2016towards} was applied for training. 

\subsection{Discussion}
\vspace{-0.3em}
We evaluated our models using PESQ~\cite{rix2001perceptual}, STOI~\cite{taal2011algorithm}, SDR~\cite{raffel2014mir_eval}, and pMOS~\cite{gamper2019intrusive} metrics, where pMOS is a neural-network-based non-intrusive MOS estimator that shows high correlations with the human MOS ratings without requiring  reference signals. Table~\ref{tab:main_results} shows the evaluation results of the seed models and their multi-task trained versions for both the simulation and real recordings. Except for the non-causal DCCRN model for the simulated test set, the seed SE models degraded the ASR performance compared with the original noisy signals. Applying the multi-task training to the seed models consistently improved the WER to a varying but significant degree. The real recording results show that the multi-task training improved the WER by 8.76\% (23.28 to 21.24), 6.30\% (20.63 to 19.33), and 11.82 \% (22.84 to 20.14) for DCUNet, DCCRN-non-causal, and DCCRN-causal, respectively. 

These WER improvements were obtained with little SE quality degradation.  For the real recordings, the pMOS values decreased only very marginally from 3.18 to 3.16, from 3.33 to 3.30, and from 3.20 to 3.17 for DCUNet, DCCRN-non-causal, and DCCRN-causal, respectively. 
If we did not interleave the SE-steps between the ASR-steps by setting the SE-step probability at 0.0, the WER for the DCCRN-causal was further improved to outperform the WER for the noisy signals. However, this also sacrificed the SE quality to some extent, reducing pMOS from 3.20 to 3.09. 

A similar trend was also observed for the
simulation set. There seems to be a trade-off between the ASR and SE quality.
An optimal operation point can be chosen by adjusting the SE-step probability based on the application needs. We further investigate this in Section 4.3.3.




\subsubsection{Impact of back-end ASR models}
\vspace{-0.3em}
Figure \ref{fig:ablation} (a) shows how the SE and ASR performance changed depending on the ASR back-end model used for the multi-task training. 
We used three Seq2Seq models with different model structures, which are denoted as ``strong'' (green), ``medium'' (yellow), and ``weak'' (blue) based on a WER ranking calculated for an ASR validation set. 
The result shows that
stronger ASR back-end models were more effective in closing the WER gap from the ``No Enhancement'' setting while preserving the SE improvement. 
Note that the ``strong'' ASR model was trained on the 75K-hour data containing various noisy signals. This suggests that it is important to use a powerful back-end model instead of a noise-sensitive model trained only on clean signals. 

\subsubsection{Impact of training data for ASR-step}\label{sec:in_out_domain_data}
\vspace{-0.3em}
Figure \ref{fig:ablation} (b) shows the impact that different ASR-step training sets had on the SE model's performance. The ``strong'' back-end model was used.  
The result shows that the models trained on 1302-hour in-domain data (yellow) and those trained on its 224-hour random subset (red) performed equally well.  
This indicates that it is sufficient to use a relatively small amount of in-domain data. 
Meanwhile, combining simulated data (green)  benefited the ASR performance compared with using only real data (red), which suggests the importance of acoustic diversity in terms of noise and reverberation conditions. 
It was also observed that 
using out-of-domain data (blue) degraded the pMOS score compared with using a similar amount of in-domain data (green) while they led to similar WERs. 

\subsubsection{Impact of SE-step probability}
\vspace{-0.3em}
We also investigated how the SE model's performance was impacted by the SE-step probability value, which controls how often the SE-step is performed in the multi-task training,
We used the ``strong'' Seq2Seq back-end model and ``224-hour in-domain with simulation'' training data for ASR-step.
Figure \ref{fig:ablation} (c) reveals the trade-off between pMOS and WER with the variation of ``SE-step probability.'' Performing the ASR-step more frequently resulted in ASR performance improvement at the expense of pMOS compared with the seed DCCRN model. 
When only the ASR-step was performed by setting the SE-step probability at 0, the WER surpassed that of ``No Enhancement'' condition, which substantially compromised the pMOS score. 
For a general use case, we would prefer a moderate SE-step probability such as $0.5$ for serving human listening and live captioning, which is shown in the last row (DCCRN-causal-MT) of Table.\ref{tab:main_results}.

\section{Conclusions}
In this paper, we proposed a multi-task training framework for deep learning-based SE models to make them more friendly to ASR systems. In the proposed approach, we first pre-trained the SE and ASR models separately. Then, we froze the ASR model parameters and started the multi-task training, where we interleaved two update steps: the SE-step and the ASR-step. The SE-step is the same as the conventional supervised SE training using noisy and clean speech pairs. For the ASR-step, we first enhanced the signals with the SE model, then fed the enhanced signals to the ASR model and backpropagated the cross-entropy loss. We controlled the frequency of each step with a probability parameter. The experimental results showed that our framework improved the ASR performance with small or little SE quality degradation. The ASR model used for the multi-task training was different from that used for evaluation, indicating that the resultant SE model can generalize to different back-end conditions. 
We also presented ablation study results that would guide the development of the proposed approach. 
In this paper, 
we opted to use a mix of public and private data to reflect the scale and complexity of real usage scenarios. We hope our work can encourage the research community to establish an open experimental platform for large-scale SE and ASR investigation.




\newpage
\bibliographystyle{IEEEtran}

\bibliography{mainbib}

\end{document}